\documentclass[aps,prd,twocolumn,groupedaddress,nofootinbib]{revtex4}
\pdfoutput=1  
\usepackage{graphicx,color}
\usepackage{appendix}
\usepackage{latexsym,amsmath,amssymb,graphicx,booktabs}
\usepackage{epsfig,latexsym}
\usepackage{hyperref}

\definecolor{MyBlue}{rgb}{0.15,0.15,0.70}

\hypersetup{
colorlinks=true,
citecolor=MyBlue,
linkcolor=MyBlue,
urlcolor=MyBlue
}

\usepackage{amssymb}
\usepackage{amsmath}
\usepackage{amsfonts}
\usepackage{upgreek}
\usepackage{latexsym}

\newcommand{\iBox}{\Box^{-1}}

\renewcommand\({\left(}
\renewcommand\){\right)}
\renewcommand\[{\left[}
\renewcommand\]{\right]}

\newcommand\n{{\mbox {\boldmath $\nabla$}}}
\newcommand{\ra}{\rightarrow}

\def\lsim{\raise 0.4ex\hbox{$<$}\kern -0.8em\lower 0.62
ex\hbox{$\sim$}}

\def\gsim{\raise 0.4ex\hbox{$>$}\kern -0.7em\lower 0.62
ex\hbox{$\sim$}}

\def\lbar{{\hbox{$\lambda$}\kern -0.7em\raise 0.6ex
\hbox{$-$}}}

\newcommand\eq[1]{eq.~(\ref{#1})}
\newcommand\eqs[2]{eqs.~(\ref{#1}) and (\ref{#2})}

\newcommand\Eqss[3]{Equations~(\ref{#1}), (\ref{#2}) and (\ref{#3})}

\newcommand\pa{\partial}
\newcommand\p{\partial}

\newcommand\ee{\end{equation}}
\newcommand\be{\begin{equation}}
\def\bea{\begin{array}}
\def\eea{\end{array}}\def\ea{\end{array}}
\newcommand\ees{\end{eqnarray}}
\newcommand\bees{\begin{eqnarray}}
\def\nn{\nonumber}

\def\a{\alpha}

\def\s{\sigma}
\def\g{\gamma}

\def\d{\delta}

\def\eps{\epsilon}

\def\dslash{\hspace{-1mm}\not{\hbox{\kern-2pt $\partial$}}}
\def\Dslash{\not{\hbox{\kern-4pt $D$}}}
\def\pslash{\not{\hbox{\kern-2.1pt $p$}}}
\def\kslash{\not{\hbox{\kern-2.3pt $k$}}}
\def\qslash{\not{\hbox{\kern-2.3pt $q$}}}

\newcommand{\vx}{{\bf x}}

\def\p1{{\bf p}_1}
\def\p2{{\bf p}_2}
\def\k1{{\bf k}_1}
\def\k2{{\bf k}_2}

\newcommand{\emn}{\eta_{\mu\nu}}

\newcommand{\eMN}{\eta^{\mu\nu}}
\newcommand{\eRS}{\eta^{\rho\sigma}}
\newcommand{\eMR}{\eta^{\mu\rho}}
\newcommand{\eNS}{\eta^{\nu\sigma}}
\newcommand{\eMS}{\eta^{\mu\sigma}}
\newcommand{\eNR}{\eta^{\nu\rho}}

\newcommand{\gmn}{g_{\mu\nu}}

\newcommand{\hmn}{h_{\mu\nu}}
\newcommand{\hrs}{h_{\rho\sigma}}

\newcommand{\bhmn}{\bar{h}_{\mu\nu}}

\newcommand{\bhMN}{\bar{h}^{\mu\nu}}

\newcommand{\parho}{\pa_{\rho}}

\newcommand{\paM}{\pa^{\mu}}
\newcommand{\paN}{\pa^{\nu}}

\newcommand{\Gmn}{G_{\mu\nu}}

\newcommand{\Tmn}{T_{\mu\nu}}

\newcommand{\TMN}{T^{\mu\nu}}

\newcommand{\dddM}{\kern 0.2em \raise 1.9ex\hbox{$...$}\kern -1.0em \hbox{$M$}}
\newcommand{\dddQ}{\kern 0.2em \raise 1.9ex\hbox{$...$}\kern -1.0em \hbox{$Q$}}
\newcommand{\dddI}{\kern 0.2em \raise 1.9ex\hbox{$...$}\kern -1.0em\hbox{$I$}}
\newcommand{\dddJ}{\kern 0.2em \raise 1.9ex\hbox{$...$}\kern-1.0em
\hbox{$J$}}
\newcommand{\dddcalJ}{\kern 0.2em \raise 1.9ex\hbox{$...$}\kern-1.0em
\hbox{${\cal J}$}}

\newcommand{\dddO}{\kern 0.2em \raise 1.9ex\hbox{$...$}\kern -1.0em
\hbox{${\cal O}$}}
\def\dddz{\raise 1.5ex\hbox{$...$}\kern -0.8em \hbox{$z$}}
\def\dddd{\raise 1.8ex\hbox{$...$}\kern -0.8em \hbox{$d$}}
\def\dddbd{\raise 1.8ex\hbox{$...$}\kern -0.8em \hbox{${\bf d}$}}
\def\ddbd{\raise 1.8ex\hbox{$..$}\kern -0.8em \hbox{${\bf d}$}}
\def\dddx{\raise 1.6ex\hbox{$...$}\kern -0.8em \hbox{$x$}}

\newcommand{\ode}{\Omega_{\rm DE}}
\newcommand{\oma}{\Omega_{M}}
\newcommand{\ora}{\Omega_{R}}

\newcommand{\ola}{\Omega_{\Lambda}}

\newcommand{\rde}{\rho_{\rm DE}}

\newcommand{\rlam}{\rho_{\Lambda}}

\begin{document}

\title{Phantom dark energy from  non-local  infrared modifications\\ \vspace{1mm} of General Relativity}

\author{Michele Maggiore}

\affiliation{D\'epartement de Physique Th\'eorique and Center for Astroparticle Physics, 
Universit\'e de Gen\`eve, 24 quai Ansermet, CH--1211 Gen\`eve 4, Switzerland}

\begin{abstract}

We discuss the cosmological consequences of a model based on a
non-local infrared modification of Einstein equations. We find that the model   
generates a dynamical dark energy that can account for the presently observed value of $\ode$, without introducing a cosmological constant. Tuning a free mass parameter $m$ to a value $m\simeq 0.67 H_0$ we reproduce the observed value $\ode\simeq 0.68$. This leaves us with no free parameter and we then get a pure prediction for the EOS parameter of dark energy. Writing
$w_{\rm DE}(a)=w_0+(1-a) w_a$,  we find
$w_0\simeq-1.04$ and $w_a\simeq -0.02$, consistent with the Planck data, and on the phantom side. We also argue that non-local equations of the type that we propose
must be understood as purely classical effective equations, such as those derived in semiclassical gravity for the in-in matrix elements of the metric. As such, any apparent ghost instability in such equations only affects the classical dynamics, but there is no propagating degree of freedom associated to the ghost, and no issue of 
ghost-induced quantum vacuum decay.

\end{abstract}

\maketitle

\section{Introduction}

In recent years there has been  an intense search for  modifications of General Relativity  (GR) that change its behavior in the far infrared, i.e. at cosmological distances, while retaining its successes at solar system scales and below. Beside an intrinsic 
field-theoretical interest, such studies are motivated by the aim of explaining the observed acceleration of the Universe. A natural way to modify GR  in the far infrared  is to introduce in the theory a mass scale $m$, of order of the present value of the Hubble parameter $H_0$. At first sight, the simplest way to introduce such a mass scale could be to build a theory where the graviton is massive. However, as is well known, 
the construction of a consistent theory with a massive graviton is quite non-trivial \cite{Fierz:1939ix,Boulware:1973my}. A significant step forward in recent years has been the construction of a ghost-free theory of massive gravity, the dRGT
theory
\cite{deRham:2010ik,deRham:2010kj} (see also \cite{deRham:2011rn,Hassan:2011hr,Hassan:2011tf,Hassan:2011ea}, and ref.~\cite{Hinterbichler:2011tt} for  a review). 
However, the present theoretical situation is not yet  satisfactory. The dRGT theory has  potential problems of acausality
over some backgrounds \cite{deRham:2011pt,Deser:2012qx,Nicolis:2009qm,Burrage:2011cr} and of strong coupling due to quantum effects \cite{Burrage:2012ja}.
Another problem is that  it is presently unclear whether a satisfactory cosmology  emerges. Spatially flat isotropic Friedman-Robertson-Walker (FRW) solutions do not even exist, and are in fact forbidden by  the same constraint that removes the ghost \cite{D'Amico:2011jj}. There are open isotropic FRW solutions, which however suffer of  strong coupling and ghost-like instabilities \cite{DeFelice:2013awa}.
Furthermore, the construction of such a  theory invokes the existence of an external reference metric, which is arguably not very natural.

In a recent paper \cite{Jaccard:2013gla} it has been proposed a different approach, that allows us to introduce a mass parameter  in GR while retaining general covariance, and does not require  an external reference metric. We found that this can be realized by introducing non-local terms. In particular, in \cite{Jaccard:2013gla} it was studied a model defined by
\be\label{final1summary}
\Gmn -m^2\(\iBox_g\Gmn\)^{\rm T}=8\pi G\,\Tmn\, ,
\ee
where the superscript T denotes the extraction of the transverse part, and $\Box_g$ is the d'Alembertian with respect to the metric $\gmn$. Its inverse $\iBox_g$ is here defined using the retarded Green's function, which ensures causality. In this paper we rather consider the model defined by
\be\label{modela2}
\Gmn -\frac{d-1}{2d}\, m^2\(\gmn \iBox_g R\)^{\rm T}=8\pi G\,\Tmn\, ,
\ee
where  $d$ is the number of spatial dimensions and the factor $(d-1)/2d$ is a convenient normalization of the mass parameter $m^2$. We will see here that the model 
(\ref{modela2}) has particularly  interesting cosmological properties (some of which are not shared by the model (\ref{final1summary}), as we will discuss in
\cite{Foffa:2013vma}).

Independently of their specific form, in \cite{Jaccard:2013gla}  the inclusion of non-local terms was considered as part of an  attempt to construct a consistent quantum field theory of massive gravity.  The purpose of this paper is two-fold. First, we will argue that  the proper interpretation of equations such as  (\ref{final1summary}) or (\ref{modela2}) is actually different, and that they should be understood as {\em classical} effective equations, obtained from the smoothing of some underlying more fundamental dynamics. Second, we will explore the cosmological consequences of 
\eq{modela2} and we will see that it gives a sensible and predictive model  for dark energy.

The paper is organized as follows. In sect.~\ref{sect:concept} we discuss conceptual issues that arise in non-local classical equations of motions such as 
\eq{modela2}. In sect.~\ref{sect:cosmo} we will examine its cosmological consequences, at the level of background evolution. 
We use the signature $\emn =(-,+,+,+)$ and units $\hbar=c=1$.

\section{Conceptual issues}\label{sect:concept}

\subsection{Absence of vDVZ discontinuity}

In order to understand the physical content of the classical theory defined by \eq{modela2},  
we begin by  studying the matter-matter interaction induced by such a modified Einstein equation.
Linearizing   over flat space and extracting the transverse part (as in app.~B of \cite{Jaccard:2013gla}) we get
\be\label{line1}
{\cal E}^{\mu\nu,\rho\sigma}\hrs
-\frac{d-1}{d}\, m^2 P^{\mu\nu}P^{\rho\sigma}
\hrs=-16\pi G\TMN\, ,
\ee
where  ${\cal E}^{\mu\nu,\rho\sigma}$  is the  Lichnerowicz operator  (conventions and definitions are as in \cite{Jaccard:2013gla}) and
\be
P^{\mu\nu}=\eMN-(\paM\paN/\Box)\, , 
\ee
where $\Box$ is the   flat-space d'Alembertian. The corresponding effective matter-matter interaction is given by 
\be
S_{\rm eff}=\int d^{d+1}x\,\hmn\TMN\, ,
\ee
 where $\hmn$ is the solution 
of \eq{line1}. To solve this equation we use the gauge invariance of the linearized theory to fix the gauge $\paM\bhmn=0$, where $\bhmn=\hmn -(1/2)h\emn$. We also use 
$\bar{h}\equiv\eMN\bhmn =-(d-1) h/2$. Then \eq{line1} becomes
\be\label{line2}
\Box\bhMN+(m^2/d) P^{\mu\nu}\bar{h}=-16\pi G\TMN\, .
\ee
Taking the trace we get 
\be
(\Box+m^2)\bar{h}=-16\pi G T\, .
\ee
We write \eq{line2} in momentum space, 
eliminate $\bar{h}$ using 
\be
\bar{h}(k)=\frac{16\pi G}{k^2-m^2}\, \tilde{T}(k)\, , 
\ee
and 
solve for $h_{\mu\nu}(k)$, obtaining
\bees
\tilde{h}_{\mu\nu}(k)&=&\frac{16\pi G}{k^2}\[
\tilde{T}_{\mu\nu}(k)-
\frac{\eMN k^2}{(d-1)(k^2-m^2)}\tilde{T}(k)\
\right.\nn\\
&&\left. 
+\frac{m^2}{d (k^2-m^2)}\(\eMN-\frac{k^{\mu}k^{\nu}}{k^2}\)
\tilde{T}(k)\]\, .
\ees
Plugging the result  into $S_{\rm eff}$ 
and using $k^{\mu}\tilde{T}_{\mu\nu}(k)=0$ to eliminate the term $k^{\mu}k^{\nu}$, we get
\be
S_{\rm eff} =16\pi G\int \frac{d^{d+1}k}{(2\pi)^{d+1}}\, \tilde{T}_{\mu\nu}(-k)\Delta^{\mu\nu\rho\sigma}(k)
\tilde{T}_{\rho\sigma}(k)\, , 
\ee
with 
\bees
\Delta^{\mu\nu \rho\s}(k)&=&\frac{1}{2k^2}\, 
\( \eMR\eNS +\eMS\eNR-\frac{2}{d-1}\eMN\eRS \) \nn\\
&&+ \frac{1}{d(d-1)}\, \frac{m^2}{k^2(-k^2+m^2)}\eMN\eRS\, .\label{Delta}
\ees
The term in the first line is the usual GR result, in generic $d$, due to the exchange of a massless graviton. The term in the second line vanishes for $m\ra 0$. Therefore this  theory has no vDVZ discontinuity and,
taking $m\sim H_0$, it smoothly reduces to GR inside the horizon. Well inside the horizon $|k^2|\gg m^2$  and the term in the second line of \eq{Delta} is of order $m^2/k^4$, compared to the massless graviton propagator which is of order $1/k^2$. Thus, at least at the level of linearized theory, well inside the horizon the predictions of this non-local theory  differ from the predictions of GR by a factor $1+{\cal O}(m^2/k^2)$. For 
$m\sim H_0$ and $|k|=(1\, {\rm a.u.})^{-1}$ (as appropriate to solar system experiments),
$m^2/k^2\sim (1\, {\rm a.u.}/H_0^{-1})^2\sim 10^{-30}$, and the predictions of the non-local theory, linearized over flat space, are indistinguishable from that of linearized GR. Further work (in progress) is need to study the behavior of perturbations both around non-trivial static solutions, as well as around the cosmological solutions that will be presented below.

\subsection{Apparent ghosts and effective classical equations}

The above computation of the matter-matter interaction stresses the purely classical nature of the derivation. One might try to be more ambitious, and  interpret directly \eq{modela2} in terms of a  quantum field theory. 
Formally the  quadratic Lagrangian corresponding to the linearized equation of motion
(\ref{line1}) is 
\be\label{Lquadr}
{\cal L}_2=\frac{1}{2}\hmn{\cal E}^{\mu\nu,\rho\sigma}\hrs
-\frac{d-1}{2d}\, m^2\hmn P^{\mu\nu}P^{\rho\sigma}\hrs\, .
\ee
Adding the usual gauge fixing term of linearized massless gravity,
${\cal L}_{\rm gf}=-(\paN\bhmn ) (\parho\bar{h}^{\rho\mu})$, and inverting the quadratic form we get the propagator $\tilde{D}^{\mu\nu \rho\s}(k)$. The explicit computation shows that, as expected, 
$\tilde{D}^{\mu\nu \rho\s}(k)=-i\Delta^{\mu\nu \rho\s}(k)$, with 
$\Delta^{\mu\nu \rho\s}(k)$ given in \eq{Delta}
(plus, as usual, terms proportional to $k^{\mu}k^{\nu}$, $k^{\rho}k^{\sigma}$ and
$k^{\mu}k^{\nu}k^{\rho}k^{\sigma}$, that give zero when contracted with a conserved energy-momentum tensor). Thus, 
the  term on the first line of \eq{Delta} gives
the usual propagator of a massless graviton, which describes only two massless states with helicities $\pm 2$. The second line gives  an extra term in 
the saturated propagator $\tilde{T}_{\mu\nu}(-k)
\tilde{D}^{\mu\nu \rho\s}(k)\tilde{T}_{\rho\sigma}(k)$, equal to
\be\label{TT}
\frac{1}{d(d-1)}\tilde{T}(-k)\[ -\frac{i}{k^2}+\frac{i}{k^2-m^2}\]
\tilde{T}(k)\, .
\ee
This term apparently describes the exchange of a healthy massless scalar plus a ghostlike massive scalar.  However, such an interpretation would not be not correct. The difficulty of promoting this classical theory to a full quantum field theory becomes apparent once one examines the $i\eps$ factor in these propagators. As emphasized in \cite{Cline:2003gs}, depending on the choice of the $i\eps$ prescription in the propagator,  a ghost either carries negative norm (and therefore ruins the probabilistic interpretation) and positive energy, or positive norm and negative energy. The choice of the $i\eps$ prescription is not specified by the Lagrangian, and is made during the quantization procedure. For a normal particle  the usual scalar propagator is $-i/(k^2+m^2-i\eps)$ (with our $(-,+,+,+)$ signature). This $i\eps$ prescription propagates positive energies forward in time. For a ghost the sign in front of the kinetic term changes, and we have in principle two options for the $i\eps$ prescription,  $i/(k^2-m^2\pm i\eps)$ (furthermore, now a positive $m^2$ gives rise to a tachyonic instability in the classical equations). The $+i\eps$  choice propagates negative energies forward in time but preserves the unitarity of the theory and the optical theorem. With the $-i\eps$  choice, in contrast, ghosts carry positive energy but negative norm, and the probabilistic interpretation of QFT is lost. This latter choice is therefore unacceptable. If in
\eq{TT}
we use the prescription that preserves positive norms, the term in brackets becomes
\be
-\frac{i}{k^2-i\eps}+\frac{i}{k^2-m^2+i\eps}
\ee
We see that now for $m=0$ these two terms no longer cancel. Thus, if one uses this prescription for the propagators, one finds  a rather bizarre situation in which at the classical level the limit $m\ra 0$ is smooth, while at the quantum level is not. The ghost is now apparently a radiative field that destabilizes the vacuum, despite the fact that, classically,  in the limit $m\ra 0$ it reduces to a non-radiative degree of freedom of GR.
If  we instead impose continuity as $m\ra 0$, we are forced to chose the $i\eps$ prescription for the ghost propagator that violates the probabilistic interpretation.\footnote{For this reason the argument in \cite{Jaccard:2013gla} that the vacuum decay amplitude induced by the  ghost is suppressed by powers of $m^2$ is unfortunately incorrect. This argument was based on the continuity of the $m\ra 0$ limit, which however only holds with the $-i\eps$ prescription that ruins the probabilistic interpretation.} None of these two options is meaningful. Indeed, 
the $m^2/\Box$ operator in eqs.~(\ref{modela2}) or (\ref{line1}) automatically comes with a retarded prescription. Since it is this term that gives rise to   the two propagators in \eq{TT}, these terms inherit the retarded prescription and cannot be promoted to Feynman propagators. They describe classical radiation effects from already existing degrees of freedom, and not new propagating degrees of freedom. 

A related observation is that a Lagrangian involving a $\iBox$ operator never gives a retarded $\iBox$ in the equations of motion, independently of the Green's function used in the definition of the $\iBox$ operator that appears in the Lagrangian. This is most easily illustrated in the case of a scalar field.
Consider for illustration a non-local term in an action of the form
\be
\int d^4x\,  \phi\iBox\phi\, , 
\ee
where $\phi$ is some scalar field, and $\iBox$ is defined with respect to some Green's function $G(x;x')$. Taking the variation with respect to $\phi(x)$ we get
\bees
&&\frac{\d}{\d\phi(x)}\int d^4x' \phi(x') (\iBox\phi )(x')\nn\\
&&=\frac{\d}{\d\phi(x)} \int d^4x' d^4x'' \phi(x') G(x';x'') \phi(x'')\nn\\
&&=\int d^4x' [G(x;x')+G(x';x)] \phi(x')\, . \label{symGreen}
\ees
We see that the variational of the action automatically symmetrizes the Green's
function \cite{Deser:2007jk,Barvinsky:2011rk,Jaccard:2013gla}. Similarly, taking the variation of  \eq{Lquadr} does not really give \eq{line1} as equation of motion:
even if  we use 
$\hmn P_{\rm ret}^{\mu\nu}P_{\rm ret}^{\rho\sigma}\hrs$ in the action, in the equation of motions we rather get $(P_{\rm ret}^{\mu\nu}P_{\rm ret}^{\rho\sigma}+
P_{\rm adv}^{\mu\nu}P_{\rm adv}^{\rho\sigma})\hrs$.
We can still use this Lagrangian as  part of a formal trick for obtaining the classical equations of motion from an action principle, in which after taking the variation we replace by hand
$\iBox_{\rm sym}\ra\iBox_{\rm ret}$. However, any direct connection to a fundamental {\em quantum} field theory  is then lost.

This suggests that we should not attempt to promote the model defined by
\eq{modela2} directly to
a full quantum field theory, but we should rather consider it just as a {\em classical} effective equation of motion. The connection to a fundamental QFT will be less direct, and we expect that it will typically involve some form of classical or quantum averaging. For instance, such effective non-local  (but causal) equations 
govern the  dynamics of the in-in matrix elements of quantum fields, such as 
$\langle 0_{\rm in}|\hat{\phi}|0_{\rm in}\rangle$ or $\langle 0_{\rm in}|\hat{g}_{\mu\nu}|0_{\rm in}\rangle$, and encode  quantum corrections  
to the classical dynamics \cite{Jordan:1986ug}.
Similar non-local equations  also emerge in a purely classical context when one separates the dynamics of a system into a long-wavelength and a short-wavelength part. One can then obtain an effective non-local equation for the long-wavelength modes by integrating out the short-wavelength modes, see e.g. \cite{Carroll:2013oxa} for a recent example in the context of cosmological perturbation theory. 
One more example comes from the standard post-Newtonian/post-Minkowskian formalisms for GW production  in GR \cite{Blanchet:2006zz,Maggiore:1900zz}.  In linearized theory the gravitational wave (GW) amplitude $\hmn$ is determined by  
$\Box\bhmn=-16\pi G\Tmn$, which in such a radiation problem is solved with the retarded Green's function,  
$\bhmn=-16\pi G\iBox_{\rm ret}\Tmn$. When the non-linearities of GR are included,
the GWs generated at some perturbative order become themselves sources for the GW generation at the next order. In the far-wave zone, this iteration gives rise to effective equations involving $\iBox_{\rm ret}$.

Trying to quantize \eq{modela2} is like trying to quantize such effective non-local equations, and makes no sense.
Simply, \eq{modela2} must be regarded as an effective {\em classical} equation and any issue of quantization, ghost, etc. 
can only be addressed in the underlying fundamental theory (see also the discussion in \cite{Creminelli:2005qk}, where it is nicely shown that the very existence of a ghost depends on the UV completion of the theory). A more extended discussion of issues related to this  ``fake" ghost will be given in \cite{Foffa:2013sma}.

\section{Cosmological evolution}\label{sect:cosmo}

\subsection{Evolution equations}

Having better understood the conceptual status of this model, we can now move to extracting its cosmological consequences.
To obtain the cosmological equations governing the background we proceed as in Sect.~8 of
 \cite{Jaccard:2013gla}. We introduce a scalar field $U$ from
$U=-\iBox R$, so   
\be\label{modela2FRW}
\Gmn +\frac{d-1}{2d}\, m^2(U\gmn)^{\rm T}=8\pi G\,\Tmn\, ,
\ee
where 
\be\label{BoxUR}
-\Box U=R\, . 
\ee
The introduction of the auxiliary field $U$ is technically convenient because it allows us to transform the original integro-differential equation into a set of differential equations.
However, it is important to observe that at the same time this procedure introduces spurious solutions. This is due to the fact that the most general solution of \eq{BoxUR}
is given by a particular solution of the inhomogeneous equation, plus the most general solution of the homogeneous equation $\Box U=0$. However, once we define the operator $\iBox$ that enters in the original non-local equation (i.e. we specify the corresponding Green's function), the initial conditions on $U$, and hence the homogeneous solution, are uniquely fixed. For instance, in a FRW metric  in $d$ spatial dimensions, $ds^2=-dt^2+a^2(t)d\vx^2$, the d'Alembertian operator on a scalar is given by $\Box f=-a^{-d}\pa_0(a^d\pa_0f)$. A possible inversion is then given 
by \cite{Deser:2007jk}
\be\label{iBoxDW}
(\iBox  R)(t)=-\int_{t_*}^{t} dt'\, \frac{1}{a^d(t')}
\int_{t_*}^{t'}dt''\, a^d(t'') R(t'')\, ,
\ee
where $t_*$ is some initial value of time (that can be taken for instance  as a value of time for which an effective description in terms of the non-local equation
(\ref{modela2}) becomes appropriate). With this definition, $U\equiv -\iBox R$ is such that $U(t_*)=0$ and $U'(t_*)=0$, so the initial conditions on $U$ are fixed once we specify what we mean by $\iBox R$. More generally, we could define $\iBox$ such that
\bees
U(t)&\equiv&  -\iBox_{\rm ret} R \\
&\equiv& U_{\rm hom}(t)+\int_{t_*}^{t} dt'\, \frac{1}{a^d(t')}
\int_{t_*}^{t'}dt''\, a^d(t'') R(t'')\, ,\nn
\ees
where $U_{\rm hom}(t)$ is a given solution of $\Box U=0$. The point is that each definition of the $\iBox$ operator, i.e. each definition of the original non-local theory, corresponds to one and only one choice of the homogeneous solution and therefore of the initial conditions for $U$. The ``free field" that satisfies the homogeneous equation $\Box U=0$ seems a propagating degree of freedom from the point of view of the local formulation (\ref{modela2FRW}),
(\ref{BoxUR}), but in fact in the original non-local theory it is not a propagating degree of freedom. Rather, each possible choice of the homogeneous solution corresponds to one definition of $\iBox$, and therefore to one specific non-local theory. 
In \cite{Foffa:2013sma} we will discuss in great detail this issue, as well as its relation with the apparent ghost degree of freedom.

Observe also that, in the local formulation  given by \eqs{modela2FRW}{BoxUR}, the theory  is invariant under the transformation
\be\label{symdegrav}
U\ra  U+c_{\Lambda}\, ,\quad
\Tmn\ra\Tmn+[\Lambda/(8\pi G)]\gmn\, ,
\ee
(while $\gmn\ra \gmn$) with $c_{\Lambda}\equiv [2d/(d-1)] (\Lambda/m^2)$,
which can be seen as a realization of the  degravitation idea~\cite{Dvali:2000xg,Dvali:2002pe,ArkaniHamed:2002fu,Dvali:2006su,Dvali:2007kt}. However, this transformation changes the boundary conditions on $U$ and therefore, from the point of view of the original non-local formulation, connects different theories, and is not a symmetry of a given non-local theory.

To write down the equations for the cosmological evolution
we now define $S_{\mu\nu}=-U\gmn$
and we split it into its transverse and longitudinal parts,
\be
S_{\mu\nu}=S^{\rm T}_{\mu\nu}+(1/2)(\n_{\mu}S_{\nu}+\n_{\nu}S_{\mu})\, . 
\ee
To determine  $S_{\mu}$  we apply $\n^{\mu}$ to both sides of this equation. 
We henceforth specialize to a flat Friedmann-Robertson-Walker (FRW) spacetime.
In FRW, the three-vector $S^i$  vanishes because there is no preferred spatial direction, while  for $S_0$ we get\footnote{Again, here we have reduced the non-local operation of taking the transverse part to a local differential equation, involving now the operator 
${\cal D}=\pa_0^2+dH\pa_0-dH^2$. Just as for the inversion of the $\iBox$ operator in
$\Box U=-R$, the initial conditions on $S_0$ are not free parameters. Rather, the definition of the non-local theory is completed once we define ${\cal D}^{-1}$, which in turn fixes the initial conditions on $S_0$, see the more extended discussion
in \cite{Foffa:2013vma}.}
\be\label{eqS0}
\ddot{S}_0+dH\dot{S}_0-dH^2S_0
=\dot{U}\, , 
\ee
In FRW, \eq{BoxUR} becomes
\be
\ddot{U}+dH\dot{U}=2d\dot{H}+d(d+1)H^2\, ,\label{BoxUFRW}\\
\ee
Finally, 
since the left-hand side of \eq{modela2} is transverse by construction, the energy-momentum tensor $\Tmn$ is covariantly conserved. Thus, to study the cosmological evolution we only need  the $(0,0)$ component of \eq{modela2}, i.e. the Friedmann equation, 
\be\label{eq2}
H^2-\frac{m^2}{d^2}(U-\dot{S}_0)=\frac{16\pi G}{d(d-1)}\rho\, .
\ee
\Eqss{eqS0}{BoxUFRW}{eq2}  give three
differential equations for the three variables $\{H(t),U(t),S_0(t)\}$.
We now take  $\rho$ to be the sum of the energy densities of matter and radiation,
$\rho =\rho_M+\rho_R$, and we set  $d=3$. We do not  include a cosmological constant   term $\rlam$. Indeed, our aim is to see whether a phenomenologically viable dark energy  model  can be obtained  from the term proportional to  $m^2$ in \eq{eq2}.
We parametrize the temporal evolution using
$x\equiv \ln a(t)$ instead of $t$,  we denote $df/dx=f'$, and we define 
\be\label{defY}
Y=U-\dot{S}_0\, . 
\ee
We also use the standard notation $h=H/H_0$, 
$\Omega_i (t)=\rho_i(t)/\rho_c(t)$ (where $i$ labels radiation, matter and dark energy), and  $\Omega_i\equiv \Omega_i (t_0)$, where $t_0$ is the present value of cosmic time.
After simple manipulations, 
 the final form of the evolution equations is as follows. The Friedmann equation reads
\be
h^2(x)=\Omega_M e^{-3x}+\Omega_R e^{-4x}+\g Y(x)
\, ,\label{hLCDM}\\
\ee
where 
\be
\g\equiv  m^2/(9H_0^2)\, . 
\ee
This shows that  there is an effective DE density
\be
\rde(t)=\rho_0\g Y(x)\, , 
\ee
where $\rho_0=3H_0^2/(8\pi G)$. The evolution of $Y(x)$ is obtained from the coupled system
\bees
&&\hspace*{-5mm}Y''+(3-\zeta)Y'-3(1+\zeta)Y=3U'-3(1+\zeta)U\, ,\label{sy1}\\
&&\hspace*{-5mm}U''+(3+\zeta)U'=6(2+\zeta)\label{sy3}\, ,\\
&&\hspace*{-5mm}\zeta(x)\equiv\frac{h'}{h}=-\, \,
\frac{3\Omega_M e^{-3x}+4\Omega_R e^{-4x}
-\g Y' }{2(\Omega_M e^{-3x}+\Omega_R e^{-4x}+\g Y)}\label{syz}\, .
\ees
We define $w_{\rm DE}$ from
\be\label{defwDE}
\dot{\rho}_{\rm DE}+3(1+w_{\rm DE})H\rho_{\rm DE}=0\, .
\ee
Using
$\dot{\rho}=H\rho'$ we see  that the equation of state (EOS) parameter of DE is
\be\label{wg2}
w_{\rm DE}(x)=-1 -\frac{Y'(x)}{3Y(x)}\, .
\ee
The same expression for $w_{\rm DE}(x)$ can  be obtained taking the trace of the $(i,j)$ component of 
\eq{modela2}. In a $d=3$ FRW space-time this gives
\be
2\dot{H}+3H^2-\frac{m^2}{3}(U-HS_0)=-8\pi G p\, ,
\ee
which can be rewritten as 
\be
2\dot{H}+3H^2=-8\pi G (p +p_{\rm DE})\, ,
\ee 
with 
\be
p_{\rm DE} = -\rho_0 \gamma (U-HS_0)\, .
\ee
From this we get 
\be
w_{\rm DE}=\frac{p_{\rm DE}}{\rho_{\rm DE}}=-\frac{U-HS_0}{U-\dot{S}_0}\, .
\ee
This can be rewritten as
\be\label{wDEdef2}
w_{\rm DE}=-1-\frac{\dot{S}_0-HS_0}{U-\dot{S}_0}\, ,
\ee
Using \eq{defY}, together with \eq{eqS0} in $d=3$, we see that \eq{wDEdef2} is equivalent to \eq{wg2}. This is of course a consequence of the fact that \eq{modela2} can be rewritten as
\be
\Gmn =8\pi G\,\(\Tmn +T_{\mu\nu}^{\rm DE}\)\, ,
\ee
where (in $d=3$)
\be
T_{\mu\nu}^{\rm DE}=\gamma\rho_0\(\gmn \iBox_g R\)^{\rm T}\, ,
\ee
and by construction $\n^{\mu}T_{\mu\nu}^{\rm DE}=0$.

\subsection{Perturbative solutions}

Before performing the numerical integration, we can get some analytic insight into the equations. In particular we can work perturbatively in $\gamma$, assuming that 
the contribution of the function $Y$ to $\zeta(x)$   is  negligible at $x$ large and negative, so that we recover standard cosmology at early times, and we then check a posteriori the self-consistency of the procedure. In this case, in each given era $\zeta(x)$ can be approximated by a constant $\zeta_0$, with $\zeta_0=-2$ in RD and $\zeta_0=-3/2$ in MD.
Then \eq{sy3} can be integrated analytically, 
\be \label{pertU}
U(x)=\frac{6(2+\zeta_0)}{3+\zeta_0}x+u_0
+u_1 e^{-(3+\zeta_0)x}\, ,
\ee
where the coefficients $u_0,u_1$ parametrize the general solution of the homogeneous equation $U''+(3+\zeta_0)U=0$. For the moment we consider the most general homogeneous solution. However, as discussed below \eq{BoxUR}, each definition of the non-local theory corresponds to one and only one choice of the homogeneous solution.
Plugging \eq{pertU} into \eq{sy1} and solving for $Y(x)$ we get
\bees
&&\hspace*{-5mm}Y(x)=-\frac{2(2+\zeta_0)\zeta_0}{(3+\zeta_0)(1+\zeta_0)}
+\frac{6(2+\zeta_0)}{3+\zeta_0} x+u_0\nn\\
&&\hspace*{-5mm}-\frac{6(2+\zeta_0) u_1 }{2\zeta_0^2+3\zeta_0-3}e^{-(3+\zeta_0)x} 
 +a_1 e^{\a_{+}x}+ a_2 e^{\a_{-}x}\, ,\label{Ypert}
\ees
where 
\be
\a_{\pm}=\frac{1}{2}\[-3+\zeta_0\pm \sqrt{21+6\zeta_0+\zeta_0^2}\]\, .
\ee
Observe that in RD $\zeta_0=-2$ and the inhomogeneous solutions for $U$ and $Y$ vanish. This is a consequence of the fact that, in RD, the Ricci scalar vanishes, so  $\Box U=0$ and the only contributions to $U$ and to $(U\gmn)^T$ come from the solutions of the homogeneous equations. The  inhomogeneous solution  is self-consistent with our perturbative approach. Indeed, in a pure RD phase it just vanishes, and in a generic epoch, as $x\ra-\infty$, $Y(x)\propto x$ so its contribution to $\zeta(x)$  is  anyhow negligible compared to the term $\oma e^{-3x}$ and $\ora e^{-4x}$ in \eq{syz}. Furthermore,  in RD, $\a_{\pm}=(-5\pm\sqrt{13})/2$ are both negative. The same happens in MD, where $\a_{\pm}=(-9\pm\sqrt{57})/4$. Therefore in RD and MD all exponentials in \eqs{pertU}{Ypert} are decaying with $x$ and, apart for the constant  mode $u_0$, in the perturbative regime the inhomogeneous solution is an attractor. If we start the evolution deep into RD, with initial conditions that are not too far from the inhomogeneous term of the perturbative solution, we will be quickly driven toward it, so within this attraction basin we can set $a_1=a_2=u_1=0$. The situation is different for $u_0$.
We see from \eq{symdegrav} that a constant shift  $U\ra U+u_0$ is equivalent to introducing a cosmological constant, with $\ola=\gamma u_0$. It is clear that, whenever one finds a  model that gives a dynamical dark energy, one can always put on top of it a constant cosmological constant term. This defines a non-minimal model with one more parameter. In this paper we focus on a minimal model in which the $\iBox$ operator is defined so that  $u_0=0$ at an initial time $t_*$ chosen deep in the RD phase, and we also  set $a_1=a_2=u_1=0$.  A  more general analysis of the dependence on the definition of the $\iBox$ and ${\cal D}^{-1}$ operators, which in the local formulation corresponds to a different choice of initial conditions, will be presented in \cite{Foffa:2013vma}.

\begin{figure}[t]
\centering
\includegraphics[width=0.85\columnwidth,height=0.50\columnwidth]{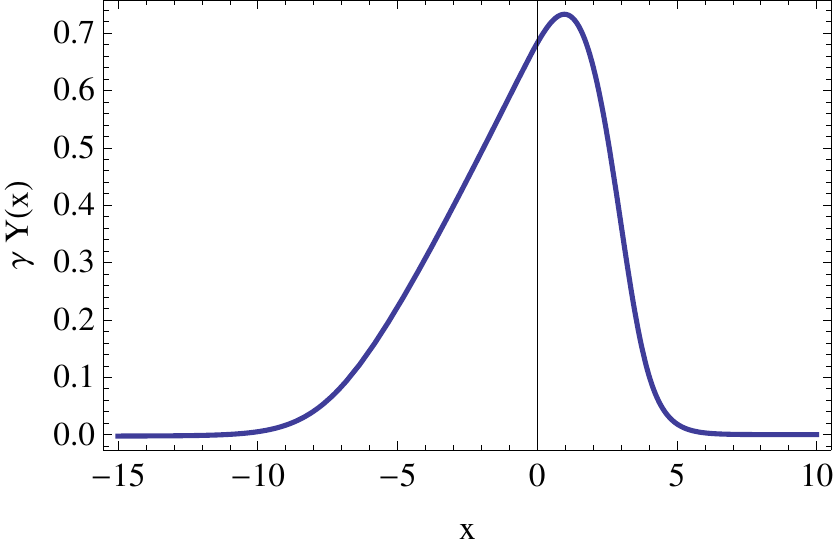}
\caption{\label{fig:DE} The function $\g Y(x)=\rde(x)/\rho_0$, against $x=\ln a$.
}
\end{figure}
\begin{figure}[t]
\centering
\includegraphics[width=0.85\columnwidth,height=0.50\columnwidth]{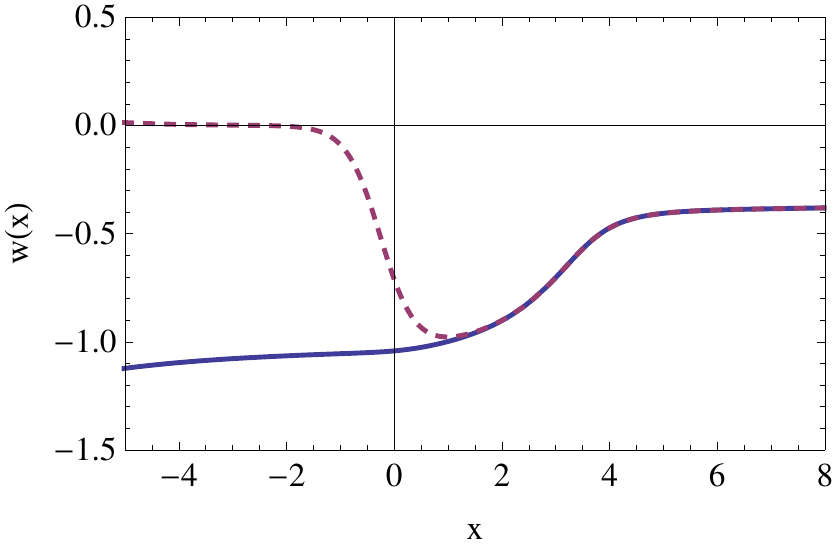}
\caption{\label{fig:w} The EOS parameter $w_{\rm DE}(x)$ (blue solid line) compared to the total EOS parameter (red dashed).
}
\end{figure}

\subsection{Full background evolution and  prediction of the DE equation of state} 

Eventually the perturbative treatment  breaks down  as we approach  the present epoch, and  we need to integrate the equations numerically.   We use the Planck best-fit  values   $\oma=0.3175$,
$\ora=4.15\times 10^{-5}h_0^{-2}$, $h_0=0.6711$ \cite{Ade:2013zuv} (and $\ola=0$). We set the initial conditions at values of $x$ deep in the RD phase  (matter-radiation equilibrium is at $x\simeq -8.1$) such that we are on the perturbative  solution found above.
The numerical solution shown in the figures is obtained setting
$\gamma =0.0504$, which corresponds to $m\simeq 0.67 H_0$. This value of $\gamma$
has been tuned so that, at $x=0$,  $\ode=1-\oma-\ora\simeq 0.6825$. 
The behavior of $\rde(x)$ (normalized to $\rho_0$) is shown in Fig.~\ref{fig:DE}. 
Having fixed $\gamma$ so that today $\ode=1-\oma-\ora$, we get a prediction (with no more free parameter) for  $\rde(x)$ at any other time. Then, from \eq{wg2}, we get a pure prediction for the dark energy  EOS parameter $w_{\rm DE}(x)$. The result is shown in fig.~\ref{fig:w}.
In MD $w_{\rm DE}(x)$ is on the phantom side and grows slowly,  and
at the present epoch is close but still slightly smaller than $-1$.
The fact that the EOS parameter is on the phantom side is generically a consequence of the fact that in our model the DE density starts from zero in RD and then grows during MD. Thus, in this regime  $\rde >0$ and $\dot{\rho}_{\rm DE}>0$, and then \eq{defwDE} implies $(1+w_{\rm DE})<0$.

For $x\gg 1$   $w_{\rm DE}$ remains negative  and $\rde (a)$ is roughly proportional to $a^{-3/2}$, so
it  still dominates over $\rho_M(x)\propto a^{-3}$. Comparing with the standard fit of the form \cite{Chevallier:2000qy,Linder:2002et}
\be
w_{\rm DE}(a)= w_0+(1-a) w_a\, ,
\ee 
(where $a(x)=e^x$) in the region $-1<x<0$, we find that
the best-fit values are
\be\label{predw0wa}
w_0 = -1.042\, ,  \qquad w_a = -0.020\, .
\ee
The  relative error between the numerical result and the fitting function is   $|\Delta w/w|\leq 2\times 10^{-4}$. 
This  non-local modification of Einstein equations therefore provides a  realization of phantom dark energy. 
It is remarkable that, without any tuning, the model produces a value of $w_0$ so close to $-1$ today, despite the fact that the time evolution of $w(x)$  is quite non-trivial, as we see from Fig.~\ref{fig:w}. 
For comparison, the   result of Planck+WP+SNLS
 for a constant $w_{\rm DE}$ (which is appropriate to our case since we predict $|w_a|\ll 1$)  is  
$ w_{\rm DE}=-1.13^{+0.13}_{-0.14}\, ,
$
at 95\% c.l. \cite{Ade:2013zuv}.

We finally observe that our prediction for $w_0$ is remarkably robust under changes of $\oma$.  If  we change the value of $\oma$ from the Planck value
$\oma=0.3175$ that we used, we should accordingly
change the value of $\ode\equiv \ode(x=0)$, so to maintain the flatness condition $\oma+\ora+\ode=1$, and this is obtained by changing $\gamma$. 
Varying $\gamma$, we find that  $\ode\simeq 0.68 (\gamma/\g_0)$, where $\g_0=0.05$, while $w_0(\gamma)\simeq w_0(\gamma_0)+5\times 10^{-3}(\g-\g_0)/\g_0$. Thus, even a revision of the value of $\oma$ at the level of 10\% would  affect our prediction for 
$w_0$  only at a level    $\Delta w_0/w_0\simeq 5\times 10^{-4}$.

In the next few years the DES survey should measure $w_0$ to an accuracy of about
$\Delta w_0\simeq 0.04$ and later EUCLID should measure it  to an accuracy
$\Delta w_0\simeq 0.01$ (and $w_a$  to an accuracy $\Delta w_a\simeq 0.1$)  \cite{Amendola:2012ys}. Such measurements will provide a stringent test of the prediction
given in \eq{predw0wa}.

\vspace{5mm}\noindent
{\bf Acknowledgements}.
We thank  Stefano Foffa, Alex Kehagias and Ermis Mitsou for very useful discussions. Our work is supported by the Fonds National Suisse.

\bibliography{myrefs_massive}

\begin{thebibliography}{36}
\expandafter\ifx\csname natexlab\endcsname\relax\def\natexlab#1{#1}\fi
\expandafter\ifx\csname bibnamefont\endcsname\relax
  \def\bibnamefont#1{#1}\fi
\expandafter\ifx\csname bibfnamefont\endcsname\relax
  \def\bibfnamefont#1{#1}\fi
\expandafter\ifx\csname citenamefont\endcsname\relax
  \def\citenamefont#1{#1}\fi
\expandafter\ifx\csname url\endcsname\relax
  \def\url#1{\texttt{#1}}\fi
\expandafter\ifx\csname urlprefix\endcsname\relax\def\urlprefix{URL }\fi
\providecommand{\bibinfo}[2]{#2}
\providecommand{\eprint}[2][]{\url{#2}}

\bibitem[{\citenamefont{Fierz and Pauli}(1939)}]{Fierz:1939ix}
\bibinfo{author}{\bibfnamefont{M.}~\bibnamefont{Fierz}} \bibnamefont{and}
  \bibinfo{author}{\bibfnamefont{W.}~\bibnamefont{Pauli}},
  \bibinfo{journal}{Proc.Roy.Soc.Lond.} \textbf{\bibinfo{volume}{A173}},
  \bibinfo{pages}{211} (\bibinfo{year}{1939}).

\bibitem[{\citenamefont{Boulware and Deser}(1972)}]{Boulware:1973my}
\bibinfo{author}{\bibfnamefont{D.}~\bibnamefont{Boulware}} \bibnamefont{and}
  \bibinfo{author}{\bibfnamefont{S.}~\bibnamefont{Deser}},
  \bibinfo{journal}{Phys.Rev.} \textbf{\bibinfo{volume}{D6}},
  \bibinfo{pages}{3368} (\bibinfo{year}{1972}).

\bibitem[{\citenamefont{de~Rham and Gabadadze}(2010)}]{deRham:2010ik}
\bibinfo{author}{\bibfnamefont{C.}~\bibnamefont{de~Rham}} \bibnamefont{and}
  \bibinfo{author}{\bibfnamefont{G.}~\bibnamefont{Gabadadze}},
  \bibinfo{journal}{Phys.Rev.} \textbf{\bibinfo{volume}{D82}},
  \bibinfo{pages}{044020} (\bibinfo{year}{2010}), \eprint{1007.0443}.

\bibitem[{\citenamefont{de~Rham
  et~al.}(2011{\natexlab{a}})\citenamefont{de~Rham, Gabadadze, and
  Tolley}}]{deRham:2010kj}
\bibinfo{author}{\bibfnamefont{C.}~\bibnamefont{de~Rham}},
  \bibinfo{author}{\bibfnamefont{G.}~\bibnamefont{Gabadadze}},
  \bibnamefont{and} \bibinfo{author}{\bibfnamefont{A.~J.}
  \bibnamefont{Tolley}}, \bibinfo{journal}{Phys.Rev.Lett.}
  \textbf{\bibinfo{volume}{106}}, \bibinfo{pages}{231101}
  (\bibinfo{year}{2011}{\natexlab{a}}), \eprint{1011.1232}.

\bibitem[{\citenamefont{de~Rham et~al.}(2012)\citenamefont{de~Rham, Gabadadze,
  and Tolley}}]{deRham:2011rn}
\bibinfo{author}{\bibfnamefont{C.}~\bibnamefont{de~Rham}},
  \bibinfo{author}{\bibfnamefont{G.}~\bibnamefont{Gabadadze}},
  \bibnamefont{and} \bibinfo{author}{\bibfnamefont{A.~J.}
  \bibnamefont{Tolley}}, \bibinfo{journal}{Phys.Lett.}
  \textbf{\bibinfo{volume}{B711}}, \bibinfo{pages}{190} (\bibinfo{year}{2012}),
  \eprint{1107.3820}.

\bibitem[{\citenamefont{Hassan and Rosen}(2012{\natexlab{a}})}]{Hassan:2011hr}
\bibinfo{author}{\bibfnamefont{S.}~\bibnamefont{Hassan}} \bibnamefont{and}
  \bibinfo{author}{\bibfnamefont{R.~A.} \bibnamefont{Rosen}},
  \bibinfo{journal}{Phys.Rev.Lett.} \textbf{\bibinfo{volume}{108}},
  \bibinfo{pages}{041101} (\bibinfo{year}{2012}{\natexlab{a}}),
  \eprint{1106.3344}.

\bibitem[{\citenamefont{Hassan et~al.}(2012)\citenamefont{Hassan, Rosen, and
  Schmidt-May}}]{Hassan:2011tf}
\bibinfo{author}{\bibfnamefont{S.}~\bibnamefont{Hassan}},
  \bibinfo{author}{\bibfnamefont{R.~A.} \bibnamefont{Rosen}}, \bibnamefont{and}
  \bibinfo{author}{\bibfnamefont{A.}~\bibnamefont{Schmidt-May}},
  \bibinfo{journal}{JHEP} \textbf{\bibinfo{volume}{1202}}, \bibinfo{pages}{026}
  (\bibinfo{year}{2012}), \eprint{1109.3230}.

\bibitem[{\citenamefont{Hassan and Rosen}(2012{\natexlab{b}})}]{Hassan:2011ea}
\bibinfo{author}{\bibfnamefont{S.}~\bibnamefont{Hassan}} \bibnamefont{and}
  \bibinfo{author}{\bibfnamefont{R.~A.} \bibnamefont{Rosen}},
  \bibinfo{journal}{JHEP} \textbf{\bibinfo{volume}{1204}}, \bibinfo{pages}{123}
  (\bibinfo{year}{2012}{\natexlab{b}}), \eprint{1111.2070}.

\bibitem[{\citenamefont{Hinterbichler}(2012)}]{Hinterbichler:2011tt}
\bibinfo{author}{\bibfnamefont{K.}~\bibnamefont{Hinterbichler}},
  \bibinfo{journal}{Rev.Mod.Phys.} \textbf{\bibinfo{volume}{84}},
  \bibinfo{pages}{671} (\bibinfo{year}{2012}), \eprint{1105.3735}.

\bibitem[{\citenamefont{de~Rham
  et~al.}(2011{\natexlab{b}})\citenamefont{de~Rham, Gabadadze, and
  Tolley}}]{deRham:2011pt}
\bibinfo{author}{\bibfnamefont{C.}~\bibnamefont{de~Rham}},
  \bibinfo{author}{\bibfnamefont{G.}~\bibnamefont{Gabadadze}},
  \bibnamefont{and} \bibinfo{author}{\bibfnamefont{A.~J.} \bibnamefont{Tolley}}
  (\bibinfo{year}{2011}{\natexlab{b}}), \eprint{1107.0710}.

\bibitem[{\citenamefont{Deser and Waldron}(2013)}]{Deser:2012qx}
\bibinfo{author}{\bibfnamefont{S.}~\bibnamefont{Deser}} \bibnamefont{and}
  \bibinfo{author}{\bibfnamefont{A.}~\bibnamefont{Waldron}},
  \bibinfo{journal}{Phys. Rev. Lett.} \textbf{\bibinfo{volume}{110}},
  \bibinfo{pages}{111101} (\bibinfo{year}{2013}), \eprint{1212.5835}.

\bibitem[{\citenamefont{Nicolis et~al.}(2010)\citenamefont{Nicolis, Rattazzi,
  and Trincherini}}]{Nicolis:2009qm}
\bibinfo{author}{\bibfnamefont{A.}~\bibnamefont{Nicolis}},
  \bibinfo{author}{\bibfnamefont{R.}~\bibnamefont{Rattazzi}}, \bibnamefont{and}
  \bibinfo{author}{\bibfnamefont{E.}~\bibnamefont{Trincherini}},
  \bibinfo{journal}{JHEP} \textbf{\bibinfo{volume}{1005}}, \bibinfo{pages}{095}
  (\bibinfo{year}{2010}), \eprint{0912.4258}.

\bibitem[{\citenamefont{Burrage et~al.}(2012)\citenamefont{Burrage, de~Rham,
  Heisenberg, and Tolley}}]{Burrage:2011cr}
\bibinfo{author}{\bibfnamefont{C.}~\bibnamefont{Burrage}},
  \bibinfo{author}{\bibfnamefont{C.}~\bibnamefont{de~Rham}},
  \bibinfo{author}{\bibfnamefont{L.}~\bibnamefont{Heisenberg}},
  \bibnamefont{and} \bibinfo{author}{\bibfnamefont{A.~J.}
  \bibnamefont{Tolley}}, \bibinfo{journal}{JCAP}
  \textbf{\bibinfo{volume}{1207}}, \bibinfo{pages}{004} (\bibinfo{year}{2012}),
  \eprint{1111.5549}.

\bibitem[{\citenamefont{Burrage et~al.}(2013)\citenamefont{Burrage, Kaloper,
  and Padilla}}]{Burrage:2012ja}
\bibinfo{author}{\bibfnamefont{C.}~\bibnamefont{Burrage}},
  \bibinfo{author}{\bibfnamefont{N.}~\bibnamefont{Kaloper}}, \bibnamefont{and}
  \bibinfo{author}{\bibfnamefont{A.}~\bibnamefont{Padilla}},
  \bibinfo{journal}{Phys.Rev.Lett.} \textbf{\bibinfo{volume}{111}},
  \bibinfo{pages}{021802} (\bibinfo{year}{2013}), \eprint{1211.6001}.

\bibitem[{\citenamefont{D'Amico et~al.}(2011)\citenamefont{D'Amico, de~Rham,
  Dubovsky, Gabadadze, Pirtskhalava, and Tolley}}]{D'Amico:2011jj}
\bibinfo{author}{\bibfnamefont{G.}~\bibnamefont{D'Amico}},
  \bibinfo{author}{\bibfnamefont{C.}~\bibnamefont{de~Rham}},
  \bibinfo{author}{\bibfnamefont{S.}~\bibnamefont{Dubovsky}},
  \bibinfo{author}{\bibfnamefont{G.}~\bibnamefont{Gabadadze}},
  \bibinfo{author}{\bibfnamefont{D.}~\bibnamefont{Pirtskhalava}},
  \bibnamefont{and} \bibinfo{author}{\bibfnamefont{A.~J.}
  \bibnamefont{Tolley}}, \bibinfo{journal}{Phys.Rev.}
  \textbf{\bibinfo{volume}{D84}}, \bibinfo{pages}{124046}
  (\bibinfo{year}{2011}), \eprint{1108.5231}.

\bibitem[{\citenamefont{De~Felice et~al.}(2013)\citenamefont{De~Felice,
  Gumrukcuoglu, Lin, and Mukohyama}}]{DeFelice:2013awa}
\bibinfo{author}{\bibfnamefont{A.}~\bibnamefont{De~Felice}},
  \bibinfo{author}{\bibfnamefont{A.~E.} \bibnamefont{Gumrukcuoglu}},
  \bibinfo{author}{\bibfnamefont{C.}~\bibnamefont{Lin}}, \bibnamefont{and}
  \bibinfo{author}{\bibfnamefont{S.}~\bibnamefont{Mukohyama}},
  \bibinfo{journal}{JCAP} \textbf{\bibinfo{volume}{1305}}, \bibinfo{pages}{035}
  (\bibinfo{year}{2013}), \eprint{1303.4154}.

\bibitem[{\citenamefont{Jaccard et~al.}(2013)\citenamefont{Jaccard, Maggiore,
  and Mitsou}}]{Jaccard:2013gla}
\bibinfo{author}{\bibfnamefont{M.}~\bibnamefont{Jaccard}},
  \bibinfo{author}{\bibfnamefont{M.}~\bibnamefont{Maggiore}}, \bibnamefont{and}
  \bibinfo{author}{\bibfnamefont{E.}~\bibnamefont{Mitsou}},
  \bibinfo{journal}{Phys.Rev.} \textbf{\bibinfo{volume}{D88}},
  \bibinfo{pages}{044033} (\bibinfo{year}{2013}), \eprint{1305.3034}.

\bibitem[{\citenamefont{Foffa et~al.}(2013{\natexlab{a}})\citenamefont{Foffa,
  Maggiore, and Mitsou}}]{Foffa:2013vma}
\bibinfo{author}{\bibfnamefont{S.}~\bibnamefont{Foffa}},
  \bibinfo{author}{\bibfnamefont{M.}~\bibnamefont{Maggiore}}, \bibnamefont{and}
  \bibinfo{author}{\bibfnamefont{E.}~\bibnamefont{Mitsou}}
  (\bibinfo{year}{2013}{\natexlab{a}}), \eprint{1311.3435}.

\bibitem[{\citenamefont{Cline et~al.}(2004)\citenamefont{Cline, Jeon, and
  Moore}}]{Cline:2003gs}
\bibinfo{author}{\bibfnamefont{J.~M.} \bibnamefont{Cline}},
  \bibinfo{author}{\bibfnamefont{S.}~\bibnamefont{Jeon}}, \bibnamefont{and}
  \bibinfo{author}{\bibfnamefont{G.~D.} \bibnamefont{Moore}},
  \bibinfo{journal}{Phys.Rev.} \textbf{\bibinfo{volume}{D70}},
  \bibinfo{pages}{043543} (\bibinfo{year}{2004}), \eprint{hep-ph/0311312}.

\bibitem[{\citenamefont{Deser and Woodard}(2007)}]{Deser:2007jk}
\bibinfo{author}{\bibfnamefont{S.}~\bibnamefont{Deser}} \bibnamefont{and}
  \bibinfo{author}{\bibfnamefont{R.}~\bibnamefont{Woodard}},
  \bibinfo{journal}{Phys.Rev.Lett.} \textbf{\bibinfo{volume}{99}},
  \bibinfo{pages}{111301} (\bibinfo{year}{2007}), \eprint{0706.2151}.

\bibitem[{\citenamefont{Barvinsky}(2012)}]{Barvinsky:2011rk}
\bibinfo{author}{\bibfnamefont{A.~O.} \bibnamefont{Barvinsky}},
  \bibinfo{journal}{Phys.Rev.} \textbf{\bibinfo{volume}{D85}},
  \bibinfo{pages}{104018} (\bibinfo{year}{2012}), \eprint{1112.4340}.

\bibitem[{\citenamefont{Jordan}(1986)}]{Jordan:1986ug}
\bibinfo{author}{\bibfnamefont{R.}~\bibnamefont{Jordan}},
  \bibinfo{journal}{Phys.Rev.} \textbf{\bibinfo{volume}{D33}},
  \bibinfo{pages}{444} (\bibinfo{year}{1986}).

\bibitem[{\citenamefont{Carroll et~al.}(2013)\citenamefont{Carroll,
  Leichenauer, and Pollack}}]{Carroll:2013oxa}
\bibinfo{author}{\bibfnamefont{S.~M.} \bibnamefont{Carroll}},
  \bibinfo{author}{\bibfnamefont{S.}~\bibnamefont{Leichenauer}},
  \bibnamefont{and} \bibinfo{author}{\bibfnamefont{J.}~\bibnamefont{Pollack}}
  (\bibinfo{year}{2013}), \eprint{1310.2920}.

\bibitem[{\citenamefont{Blanchet}(2006)}]{Blanchet:2006zz}
\bibinfo{author}{\bibfnamefont{L.}~\bibnamefont{Blanchet}},
  \bibinfo{journal}{Living Rev.Rel.} \textbf{\bibinfo{volume}{9}},
  \bibinfo{pages}{4} (\bibinfo{year}{2006}).

\bibitem[{\citenamefont{Maggiore}(2007)}]{Maggiore:1900zz}
\bibinfo{author}{\bibfnamefont{M.}~\bibnamefont{Maggiore}},
  \emph{\bibinfo{title}{{Gravitational Waves. Vol. 1. Theory and Experiments}}}
  (\bibinfo{publisher}{Oxford University Press, 574 p}, \bibinfo{year}{2007}).

\bibitem[{\citenamefont{Creminelli et~al.}(2005)\citenamefont{Creminelli,
  Nicolis, Papucci, and Trincherini}}]{Creminelli:2005qk}
\bibinfo{author}{\bibfnamefont{P.}~\bibnamefont{Creminelli}},
  \bibinfo{author}{\bibfnamefont{A.}~\bibnamefont{Nicolis}},
  \bibinfo{author}{\bibfnamefont{M.}~\bibnamefont{Papucci}}, \bibnamefont{and}
  \bibinfo{author}{\bibfnamefont{E.}~\bibnamefont{Trincherini}},
  \bibinfo{journal}{JHEP} \textbf{\bibinfo{volume}{0509}}, \bibinfo{pages}{003}
  (\bibinfo{year}{2005}), \eprint{hep-th/0505147}.

\bibitem[{\citenamefont{Foffa et~al.}(2013{\natexlab{b}})\citenamefont{Foffa,
  Maggiore, and Mitsou}}]{Foffa:2013sma}
\bibinfo{author}{\bibfnamefont{S.}~\bibnamefont{Foffa}},
  \bibinfo{author}{\bibfnamefont{M.}~\bibnamefont{Maggiore}}, \bibnamefont{and}
  \bibinfo{author}{\bibfnamefont{E.}~\bibnamefont{Mitsou}}
  (\bibinfo{year}{2013}{\natexlab{b}}), \eprint{1311.3421}.

\bibitem[{\citenamefont{Dvali and Gabadadze}(2001)}]{Dvali:2000xg}
\bibinfo{author}{\bibfnamefont{G.}~\bibnamefont{Dvali}} \bibnamefont{and}
  \bibinfo{author}{\bibfnamefont{G.}~\bibnamefont{Gabadadze}},
  \bibinfo{journal}{Phys.Rev.} \textbf{\bibinfo{volume}{D63}},
  \bibinfo{pages}{065007} (\bibinfo{year}{2001}), \eprint{hep-th/0008054}.

\bibitem[{\citenamefont{Dvali et~al.}(2003)\citenamefont{Dvali, Gabadadze, and
  Shifman}}]{Dvali:2002pe}
\bibinfo{author}{\bibfnamefont{G.}~\bibnamefont{Dvali}},
  \bibinfo{author}{\bibfnamefont{G.}~\bibnamefont{Gabadadze}},
  \bibnamefont{and} \bibinfo{author}{\bibfnamefont{M.}~\bibnamefont{Shifman}},
  \bibinfo{journal}{Phys.Rev.} \textbf{\bibinfo{volume}{D67}},
  \bibinfo{pages}{044020} (\bibinfo{year}{2003}), \eprint{hep-th/0202174}.

\bibitem[{\citenamefont{Arkani-Hamed et~al.}(2002)\citenamefont{Arkani-Hamed,
  Dimopoulos, Dvali, and Gabadadze}}]{ArkaniHamed:2002fu}
\bibinfo{author}{\bibfnamefont{N.}~\bibnamefont{Arkani-Hamed}},
  \bibinfo{author}{\bibfnamefont{S.}~\bibnamefont{Dimopoulos}},
  \bibinfo{author}{\bibfnamefont{G.}~\bibnamefont{Dvali}}, \bibnamefont{and}
  \bibinfo{author}{\bibfnamefont{G.}~\bibnamefont{Gabadadze}}
  (\bibinfo{year}{2002}), \eprint{hep-th/0209227}.

\bibitem[{\citenamefont{Dvali}(2006)}]{Dvali:2006su}
\bibinfo{author}{\bibfnamefont{G.}~\bibnamefont{Dvali}}, \bibinfo{journal}{New
  J.Phys.} \textbf{\bibinfo{volume}{8}}, \bibinfo{pages}{326}
  (\bibinfo{year}{2006}), \eprint{hep-th/0610013}.

\bibitem[{\citenamefont{Dvali et~al.}(2007)\citenamefont{Dvali, Hofmann, and
  Khoury}}]{Dvali:2007kt}
\bibinfo{author}{\bibfnamefont{G.}~\bibnamefont{Dvali}},
  \bibinfo{author}{\bibfnamefont{S.}~\bibnamefont{Hofmann}}, \bibnamefont{and}
  \bibinfo{author}{\bibfnamefont{J.}~\bibnamefont{Khoury}},
  \bibinfo{journal}{Phys.Rev.} \textbf{\bibinfo{volume}{D76}},
  \bibinfo{pages}{084006} (\bibinfo{year}{2007}), \eprint{hep-th/0703027}.

\bibitem[{\citenamefont{Ade et~al.}(2013)}]{Ade:2013zuv}
\bibinfo{author}{\bibfnamefont{P.}~\bibnamefont{Ade}} \bibnamefont{et~al.}
  (\bibinfo{collaboration}{Planck Collaboration}) (\bibinfo{year}{2013}),
  \eprint{1303.5076}.

\bibitem[{\citenamefont{Chevallier and Polarski}(2001)}]{Chevallier:2000qy}
\bibinfo{author}{\bibfnamefont{M.}~\bibnamefont{Chevallier}} \bibnamefont{and}
  \bibinfo{author}{\bibfnamefont{D.}~\bibnamefont{Polarski}},
  \bibinfo{journal}{Int.J.Mod.Phys.} \textbf{\bibinfo{volume}{D10}},
  \bibinfo{pages}{213} (\bibinfo{year}{2001}), \eprint{gr-qc/0009008}.

\bibitem[{\citenamefont{Linder}(2003)}]{Linder:2002et}
\bibinfo{author}{\bibfnamefont{E.~V.} \bibnamefont{Linder}},
  \bibinfo{journal}{Phys.Rev.Lett.} \textbf{\bibinfo{volume}{90}},
  \bibinfo{pages}{091301} (\bibinfo{year}{2003}), \eprint{astro-ph/0208512}.

\bibitem[{\citenamefont{Amendola et~al.}(2013)}]{Amendola:2012ys}
\bibinfo{author}{\bibfnamefont{L.}~\bibnamefont{Amendola}} \bibnamefont{et~al.}
  (\bibinfo{collaboration}{Euclid Theory Working Group}),
  \bibinfo{journal}{Living Rev.Rel.} \textbf{\bibinfo{volume}{16}},
  \bibinfo{pages}{6} (\bibinfo{year}{2013}), \eprint{1206.1225}.

\end{thebibliography}
\end{document}